\documentclass{jaa}
\usepackage{natbib}
\setcitestyle{aysep={}, yysep={;}}

\usepackage{graphicx}
\usepackage{txfonts}
\usepackage{hyperref}
\hypersetup{colorlinks=true, linkcolor=blue, filecolor=magenta, urlcolor=cyan, citecolor=blue}
\usepackage{parskip}
\usepackage{rotating}
\usepackage{float}
\usepackage{tabularx}
\usepackage{longtable}
\usepackage{rotating}
\usepackage{xcolor,graphicx,subfig,lipsum}
\usepackage[switch]{lineno}

\begin{document}\sloppy

\title{A kinematic and structural study of young open clusters in the Milky Way Galaxy using \textit{Gaia} DR3 catalogue}

\author{Harsha K H\textsuperscript{1,2,*}, Annapurni Subramaniam\textsuperscript{2}, S. R. Dhanush \textsuperscript{2,3}, and Hariharan D S \textsuperscript{4}}
\affilOne{\textsuperscript{1}Indian Institute of Science Education and Research, Tirupati, 517619, India\\}
\affilTwo{\textsuperscript{2} Indian Institute of Astrophysics, 2nd Block Koramangala, Banglore 560034, India \\}
\affilThree{\textsuperscript{3} Pondicherry University, R.V. Nagar, Kalapet, Puducherry 605014, India\\}
\affilFour{\textsuperscript{4} Center for Astrophysics Research (CAR), University of Hertfordshire, UK}

\twocolumn[{

\maketitle

\corres{harshakh99@gmail.com}

\msinfo{14 April 2025}{24 July 2025}

\begin{abstract}
We aim to identify the cluster members, estimate cluster properties, study the dynamical state of the clusters as a function of mass, trace the existence of dynamical effects in massive stars, and check for spatial patterns of members in young clusters. We studied 14 young open clusters located within 1 kpc using the data from \textit{Gaia} DR3 with the membership estimated using the GMM method. The cluster parameters such as age, distance, metallicity, and extinction were estimated by fitting PARSEC isochrones to the CMDs. These clusters are found to have ages between 6--90 Myr, located between 334 -- 910 pc, covering a mass range, of 0.13 to 13.77 M$_\odot$. In five of these clusters, stars from F to M spectral type show increasing velocity dispersion, a signature for dynamical relaxation. We detect high proper motion for B and A-type stars, possible walkaway stars in the other five clusters, Alessi Teutsch 5, ASCC 16, ASCC 21, IC 2395, and NGC 6405. We demonstrate the existence of mass-dependent velocity dispersion in young clusters suggestive of dynamical relaxation. The typical range of transverse velocity dispersion is found to be 0.40 - 0.70 km s$^{-1}$ for young clusters. 

\end{abstract}


}]


\doinum{}
\artcitid{\#\#\#\#}
\volnum{000}
\year{0000}
\pgrange{1--}
\setcounter{page}{1}
\lp{21}

\section{Introduction}
Stars form in molecular clouds, mostly in groups or clusters \citep{lada}. Star clusters have always served as an open laboratory for various stellar evolutionary studies \citep{CG}. As star clusters are the fundamental building blocks of our Galaxy \citep{Kharchenko}, studying them can help us understand the evolution and structure of the galaxies that comprise the universe \citep{CG}.

Star clusters comprising population I stars in our Galaxy are known as Open Clusters (OCs). These are loosely bound by gravitation and go through various dynamical processes such as tidal perturbation due to the Galactic disk, internal interaction between members, and feedback from stellar evolution leading to the evaporation in 100 Myr \citep{Wielen1971}. \cite{Park-SM}, suggested from their numerical simulation that star clusters are rapidly destroyed in the inner Galactic disk, where there is a strong effect of the tidal field. \cite{Bergond-leon}, have explained the effect of time-varying tidal fields during disc shocks in the dynamical evolution of OCs. From their simulation, they suggested that about 20\% of mass is lost during this process, and the destruction time scale is around 600 Myr for the OCs. One of the major assumptions in the simulation is the internal velocity dispersion in OCs as well as the dynamical state of the cluster members. It is therefore important to derive the observed spatial and kinematical properties of young surviving open clusters to provide realistic parameters for simulations. 

The clusters go through dynamical relaxation over a time scale, leading to the segregation of massive stars towards the center and energy equipartition. The relaxation may lead to more massive stars having closer orbits and less massive stars with extended orbits with respect to the center of the cluster \citep{RamSagar1989}. \cite{Lada-Margulis-Dearborn1984} and \cite{ Margulis-Lada1984} studied the dynamical evolution of young open clusters with about 100 members and found that, as a result of two body interactions, mass segregation may occur after 2--3 Myr. This study suggested that less massive stars relax relatively quickly, leading to mass segregation. The presence of mass segregation or dynamical relaxation can be traced by detecting mass-dependent velocity dispersion in young clusters. This was attempted long back by \cite{RamSagar1989}.

Earlier, \cite{McNamara1986}, studied the internal kinematics of M35 OC, suggesting that the velocity-mass relation may not be satisfied by high-mass stars (1.2 -- 5.6 M$_\odot$). Later, \cite{RamSagar1989} agreed with this study from their findings that the equipartition of energy may not be valid for all the clusters. They also suggested that the velocity-mass relation can be affected by the galactic field, dynamical evolution, and initial star-forming conditions. Recently, \cite{Bhattacharya-Kaushar2022} studied the tidal tails and mass segregation in OCs using \textit{Gaia} EDR3. They found that mass segregation is mostly observed in older OCs.

Some of the high-mass stars, runaway stars, in the clusters, have unusually high velocity, and most of such are seen in the dense and low-mass clusters \citep{Fujii-Zwart}. The term 'Runaway Stars' was first coined by \cite{Blaauw1961} to represent high-mass stars that have unusually high velocity ($>$ 30 km s$^{-1}$). This high velocity can be due to dynamical ejection or binary interactions \citep{Poveda1967, Leonard1991, Allen-Kinman2004}. Massive stars play a key role in enriching the interstellar medium (ISM); therefore, runaway stars contribute to enriching the regions of the ISM far away from their birthplace. There could be many unbound massive stars that are slower and hence not included in the runaway stars. Many suggested there can be unbound massive stars with velocity $<$ 30 km s$^{-1}$ \citep{deMink2012, Perets-Subr2012, Eldridge2011, Renzo2019} and such stars are termed "walkaway stars" \citep{deMink2012}. With the availability of large data with precise kinematic information, studies are growing to identify such stars and their origin. 

There are four main objectives for this study. Firstly, to identify the cluster members and precise cluster properties, which can be used as probes to explore star formation and dynamics. Next, to study the dynamical state of the young clusters as a function of mass using transverse velocity dispersion. Then, to trace the existence of dynamical effects in massive stars and the existence of unbound stars such as walkaway stars. Finally, to trace the imprints and patterns of star formation in OCs. Now, with the availability of more precise and accurate astrometric and photometric data from \textit{Gaia} DR3, it is possible to derive highly reliable membership information and cluster parameters that are vital to derive the dynamical state of OCs. In this work, we studied the kinematic and structural properties of 14 OCs. We selected the target OCs by filtering the clusters that are less than 100 Myr of age and within 1 kpc distance from the Cantat-Gaudin catalog \citep{CG}. Choosing OCs within 1 kpc allows us to reduce error in parameter estimation and to detect low-mass members. 

The paper is arranged as follows: In Section \ref{sec:data}, we describe the data and methods used for the purpose of analysis. In Section \ref{sec:results and discussion}, we present the results obtained from the analysis and discuss our main findings. Finally, we conclude our paper with Section \ref{sec:Conclusion}.

\section{Data, Methods, and Analysis} \label{sec:data}

\textit{Gaia} mission surveys a large area of space (including the Milky way galaxy and a significant amount of extragalactic objects) and collects astrometric, photometric, and spectrometric data with high precision for over 1.8 billion stars \citep{Gaia-dr3}. A total of 4 data releases, DR1, DR2, EDR3, and DR3, have been released, with the recent data release (\textit{Gaia} DR3) having high precision proper motion (PM) values \cite{Gaia-dr3}. The \textit{Gaia} DR3 data for 14 clusters were downloaded from the \textit{Gaia} archive with a radius of r50$\times$2 from its central coordinate, where r50 is the radius containing half of the cluster members, obtained from \cite{CG}, Table \ref{table_apx}. In the case of Alessi Teutsch 5, we have opted for a radius smaller than that mentioned in \cite{CG} due to the presence of another cluster closer to it. The cluster parameters required for the analysis were also taken from \cite{CG}.
 
\subsection{\textbf{Data filtering}} \label{sec:fltr}
We needed to remove the obvious field stars before using the clustering algorithm. We considered sources with a parallax cut-off of $|\varpi - \frac{1}{D}| < (3 \times \varpi_{err})$, where $D$ is distance and $\varpi_{err}$ is the error in parallax. This condition removes field stars in the line of sight, \cite{E.Vasiliev}. The following filtering of the data was done following \cite{E.Vasiliev}, and \cite{kerr}:
\begin{enumerate}
\item $AEN < 1.0$ \label{eq:fltr_aen}
\item $RUWE < 1.2$ \label{eq:fltr_ruwe}
\item $(PMRA^{2} + PMDEC{^2})  <  (30.0)^{2}$ \label{eq:fltr_pmradec}
\item $P_{bprp}(excess) < 1.3 + 0.037 \times (bprp)^{2}$ \label{eq:fltr_bprpexcess}
\item $\frac{\varpi}{\varpi_{err}} > 5$ \label{eq:fltr_plx2}
\end{enumerate}
 
Where $AEN$ is the Astrometric Excess Noise, representing the excess noise in the sources arising from modeling errors and unresolved binaries, and $RUWE$ is the Renormalized Unit Weight error; a higher value of this indicates the presence of non-single stars, such as unresolved binaries and clumped-up sources. Thus, the quality filter, conditions \ref{eq:fltr_aen} and \ref{eq:fltr_ruwe}, helps to eliminate unreliable astrometric sources. Also, sources with very large PM are removed using the condition \ref{eq:fltr_pmradec}, where PMRA is the PM in the right ascension (RA) direction, and PMDEC is the PM in the declination (DEC) direction. 

$P_{bprp}(excess)$ is the photometric excess factor of a source and $bprp$ is the $G_{bp}- G_{rp}$ color. The $P_{bprp}(excess)$ measures the excess flux in the integrated BP and RP integrated photometry. The condition \ref{eq:fltr_bprpexcess} removes the faint sources that are affected by a brighter source nearby, ie., the color of the fainter sources appears to be brighter than it actually is. An additional constraint in parallax, condition \ref{eq:fltr_plx2}, is also used to remove the sources with higher errors in their parallax measurement.

\subsection{\textbf{Identification of cluster members using GMM}} \label{sec:GMM}
The crucial part of the study was to identify the cluster members. We have used the Gaussian Mixture Model (GMM) algorithm, \cite{E.Vasiliev}, for the probabilistic identification of the cluster members. The Python algorithm for GMM is available on GitHub, \href{https://github.com/GalacticDynamics-Oxford/GaiaTools/blob/master/run_fit.py}{GaiaTools}.

We used the PM information obtained from \textit{Gaia} DR3 for the sources that pass through the filtering conditions mentioned in Section \ref{sec:fltr} to identify the cluster members using GMM. The GMM assumes several Gaussian distributions to each data group within the given data. Initial conditions such as PM and velocity dispersion information obtained from the literature surveys were used to initiate the algorithm to find the best-fit Gaussian parameter using the expectation-maximization algorithm. Gaussian distribution is assumed for cluster members and field stars, and the best-fit parameters are used to find the membership probability of the individual sources. 

From this analysis, we obtained revised mean values for PMRA and PMDEC, as well as their corresponding errors for a cluster. Also, the membership probability of all sources was obtained. For further analysis, we have chosen sources with a probability $\ge$90$\%$ as the cluster members.

Even after the probability cut off, the cluster members include fainter stars that are likely to have larger uncertainties in PM and parallax \citep{CG}. Hence, we removed the sources with their G magnitude $\ge$ 19. Also, an additional filtering condition is used by defining a parameter PMR$_0$, by which the PM of the sources varies with respect to the PM from the literature survey, \cite{vikranth}.                                         
{\setlength{\abovedisplayskip}{6pt}
 \setlength{\belowdisplayskip}{6pt}
\begin{equation}
\resizebox{0.9\columnwidth}{!}{$
    \text{PMR}_0 = \sqrt{ (\text{PMRA}_s - \text{PMRA}_i)^2 + (\text{PMDEC}_s - \text{PMDEC}_i)^2 }
    $}
\end{equation}
}

Where $\text{PMRA}_s$ and $\text{PMRA}_i$ are the PM along RA for a source and the intrinsic PMRA of the cluster, respectively. And $\text{PMDEC}_s$ and $\text{PMDEC}_i$ are the PM along DEC for a source and intrinsic PMDEC for the cluster, respectively. And only sources with $\text{PMR}_0$ $<$ 2 are included for further analysis.

We have also made use of the available radial velocity measurements for the cluster members from \textit{Gaia} DR3. We used these estimations to eliminate sources having a deviation greater than $3\sigma$ from the mean radial velocity for further study.

\subsection{\textbf{Isochrone fitting}} \label{sec:iso_fit}
 
The \textit{Gaia} colour-magnitude diagram (CMD) of the clusters can be compared with isochrones to find the cluster parameters such as age, distance, metallicity, and extinction. For that, we used the PARSEC isochrone models \citep{Bressan, Chen2014, Chen2015}. The required isochrone models were downloaded from the CMD \href{http://stev.oapd.inaf.it/cgi-bin/cmd}{3.7} web interface. The ranges for $log_{10}(age)$ are set between 6.0 and 8.0 with a time step of 0.01 and metallicity ([$M/H$]) between $-$0.2  to 0.2  with a time step of 0.02. We used a combination of least-square minimization and the Markov Chain Monte Carlo (MCMC) method to fit the isochrone to the cluster CMDs, as described below. 

\subsubsection{\textbf{Least-square minimization technique}} \label{sec:chi sqr mini}

The least-square minimization was used to find the prior estimates of the parameters such as age, distance, metallicity, and extinction for MCMC sampling. We used the distance moduli equation to estimate the distance to the cluster and extinction in G mag.
{\setlength{\abovedisplayskip}{6pt}
 \setlength{\belowdisplayskip}{6pt}
\begin{equation}
M_G =  m_G - \left[5 \times \log \left(\frac{D}{10}\right)\right] - A_G
\end{equation}
}
{\setlength{\abovedisplayskip}{6pt}
 \setlength{\belowdisplayskip}{6pt}
\begin{equation}
M(bp-rp) = (bp-rp) - 0.537* A_G 
\end{equation}
}

where $M_G$ is the absolute G magnitude (G mag), $m_G$ is the apparent G mag, $D$ is the distance to the cluster, $A_G$ is the extinction in G mag, $M(bp-rp)$ is the dereddened $bp-rp$ color, and $bp-rp$ is the color without reddening correction. The relation between reddening correction for color and extinction in G mag was obtained from the CMD \href{http://stev.oapd.inaf.it/cgi-bin/cmd}{3.7} website, \cite{ref:cardelli} and \cite{O'Donnell}.

The least-square minimization algorithm finds the theoretical model that shows a close match with cluster members in their CMD. The $log_{10}(age)$ was varied by $\pm$ 0.5 with a step size of 0.05 and distance by $\pm$ 10 pc with a step size of 1pc from the $log_{10}(age)$ and distance obtained from the literature survey, respectively. Metallicity ([M/H]) was varied between -0.2 dex and 0.2 dex with a step size of 0.1 dex, and $A_G$ was varied between 0 and 0.832 with a step size of 0.1. 
{\setlength{\abovedisplayskip}{6pt}
 \setlength{\belowdisplayskip}{6pt}
\begin{equation}
S = S_{mag} + S_{color} \label{eq:chi2}
\end{equation}
}
Where,
{\setlength{\abovedisplayskip}{6pt}
 \setlength{\belowdisplayskip}{6pt}
\begin{equation}
S_{mag} = (M_{G,i} - M_{G,j})^2 
\label{eq:chi_mag}
\end{equation}
}
{\setlength{\abovedisplayskip}{6pt}
 \setlength{\belowdisplayskip}{6pt}
\begin{equation}
S_{color} = \left[M(bp-rp)_i - M(bp-rp)_j\right]^2 \label{eq:chi_color}                 
\end{equation}
}

Where $M_{G,i}$ and $ M_{G,j}$ represent the absolute G mag of cluster members and the isochrone model, respectively, and $M(bp-rp)_i$ and $M(bp-rp)_j$ represent the reddening corrected bp-rp color of cluster members and the isochrone model, respectively. 
By varying parameters as mentioned above, we found the deviation ($S$) of each model from our cluster members using the above equation. Then, we also removed the data points having a higher deviation from the CMD by using 1$\sigma$ clipping and chose only those points that passed the sigma clipping for fitting. Next, we found the mean deviation of these chosen points. The prior estimates were found by finding the model with the least mean deviation. These estimated parameters will provide us with the prior estimates for age, metallicity, distance, and extinction in the G mag of the desired cluster.

\subsubsection{\textbf{MCMC technique}} \label{sec:mcmc} 
The prior estimates obtained from the previous step are then used as the initial guesses for each cluster to run the MCMC, \cite{Hogg-Mackey}. We have used the emcee Python library, \cite{Foreman-Mackey}, for this purpose. For a better MCMC performance, we required isochrone models with better resolution. Since the step size of the models generated from PARSEC is smaller, we defined a function that could generate isochrones for any given random values of $log_{10}(age)$, [$M/H$], distance, and extinction. This synthetic generation of isochrones was achieved by a series of linear interpolations of the downloaded theoretical isochrones from PARSEC. Firstly the interpolation was done with respect to their age and then with respect to their metallicity to obtain the desired isochrone. Also, the time step mentioned in section \ref{sec:chi sqr mini} for $log_{10}(age)$ is increased from 0.05 dex to 0.01dex and that of [$M/H$] from 0.1 to 0.02. This ensures that the generated synthetic isochrones are more precise.

We defined the likelihood function as,                        
{\setlength{\abovedisplayskip}{6pt}
 \setlength{\belowdisplayskip}{6pt}
\begin{equation}
\resizebox{0.92\columnwidth}{!}{$
\ln\:L = - \frac{1}{2} \sum \left\{\chi_{\text{mag}}^2 + \chi_{\text{color}}^2 + \log(m_{Gerr,i}) + log[(bp-rp)_{err,i} \right\}
$}
\label{eq:ln_l}
\end{equation}

Where,
{\setlength{\abovedisplayskip}{6pt}
 \setlength{\belowdisplayskip}{6pt}
\begin{equation}
\chi_{mag}^2 = \frac{(m_{G,i} - m_{G,j})^2}{m_{Gerr,i}}
\label{eq:mcmc_chi_mag}
\end{equation}
}
{\setlength{\abovedisplayskip}{6pt}
 \setlength{\belowdisplayskip}{6pt}
\begin{equation}
\chi_{color}^2 = \frac{\left[(bp-rp)_i - (bp-rp)_j\right]^2}{(bp-rp)_{err,i}}
\label{eq:mcmc_chi_color}
\end{equation}
}

Where $m_{G,i}$ and $m_{G,j}$ represent the apparent G mag of cluster members and the isochrone model, respectively, and $(bp-rp)_i$ and $(bp-rp)_j$ represent the apparent (bp-rp) color of cluster members and isochrone model, respectively. While $m_{Gerr,i}$ and $(bp-rp)_{err,i}$ are the errors in the G mag and (bp-rp) color of the cluster members, respectively.
 
Since there is a spread of cluster members in the CMD, it is challenging to implement MCMC. Hence, we used fiducial points instead of the full CMD in the cluster sequence. This fiducial point is obtained by binning the cluster members with respect to G mag with a bin size of 0.1 magnitudes. Then, we removed the highly deviated data points in each bin by giving a sigma clipping of 3$\sigma$ in the (bp-rp) color. Next, we found the mean G mag and (bp-rp) color and their corresponding error using the clipped data points. The CMD obtained by plotting the mean G mag and (bp-rp) is the fiducial sequence for our cluster members. This is then used to fit the theoretical isochrones. Uniform priors were chosen such that they centered around the parameter values obtained from \ref{sec:chi sqr mini}.

Then, by giving initial guesses as prior estimates from section \ref{sec:chi sqr mini}, we ran MCMC with 72 walkers and a step size of 500 steps. We set the initial 250 steps as the burn-in phase and found all clusters to have converging parameters afterward and then found the 16th, 50th, and 84th percentile. We took the 50th percentile as the best-fit parameter for each parameter space. The difference between the 16th -- 50th percentiles and the 50th -- 84th percentiles are taken as the errors for these best-fit parameters.

\subsection{\textbf{Velocity estimation}} \label{sec:vel}

The transverse velocity, which is the velocity along the perpendicular component to our line of sight of a star, can be determined from the PM of a star.
{\setlength{\abovedisplayskip}{6pt}
 \setlength{\belowdisplayskip}{6pt}
\begin{equation}
V_{T} = 4.74 (\mu D)
\end{equation}
}
Where $V_{T}$ is the transverse velocity in km s$^{-1}$, $\mu$ is the PM in milliarcseconds per year (mas yr$^{-1}$), and $D$ is the distance from the sun in kiloparsec (kpc) of a star. By substituting for the distance obtained from the MCMC method, We found the mean transverse velocity for each cluster. Its dispersion was then found by taking its standard deviation.

We found the mean radial velocity of clusters using the radial velocity measurements of the members from \textit{Gaia} DR3 data.

Space velocity is the total velocity experienced by a cluster in the 3D space. It is obtained by considering the mean radial velocity and the mean transverse velocity. 
{\setlength{\abovedisplayskip}{6pt}
 \setlength{\belowdisplayskip}{6pt}
\begin{equation}
V = \sqrt{V_{T} ^2 + V_{R}^2}
\end{equation}
}
Where $V$ is the space velocity, $V_{T}$ is the transverse velocity, and $V_{R}$ is the radial velocity.

\subsection{\textbf{Mass estimation}} \label{sec:mass}
Photometric masses for the stars were available in the PARSEC isochrone data \citep{Bressan, Chen2014, Chen2015}. We used a similar interpolation method mentioned in section \ref{sec:mcmc} to obtain the mass sequence for the best-fit isochrone. Which was then further interpolated with respect to the G mag of the cluster members to obtain the mass of individual stars within the cluster.

\section{Results and Discussion} \label{sec:results and discussion}

\subsection{Membership information}

\begin{figure*}
    \centering
    \includegraphics[width=0.95\linewidth]{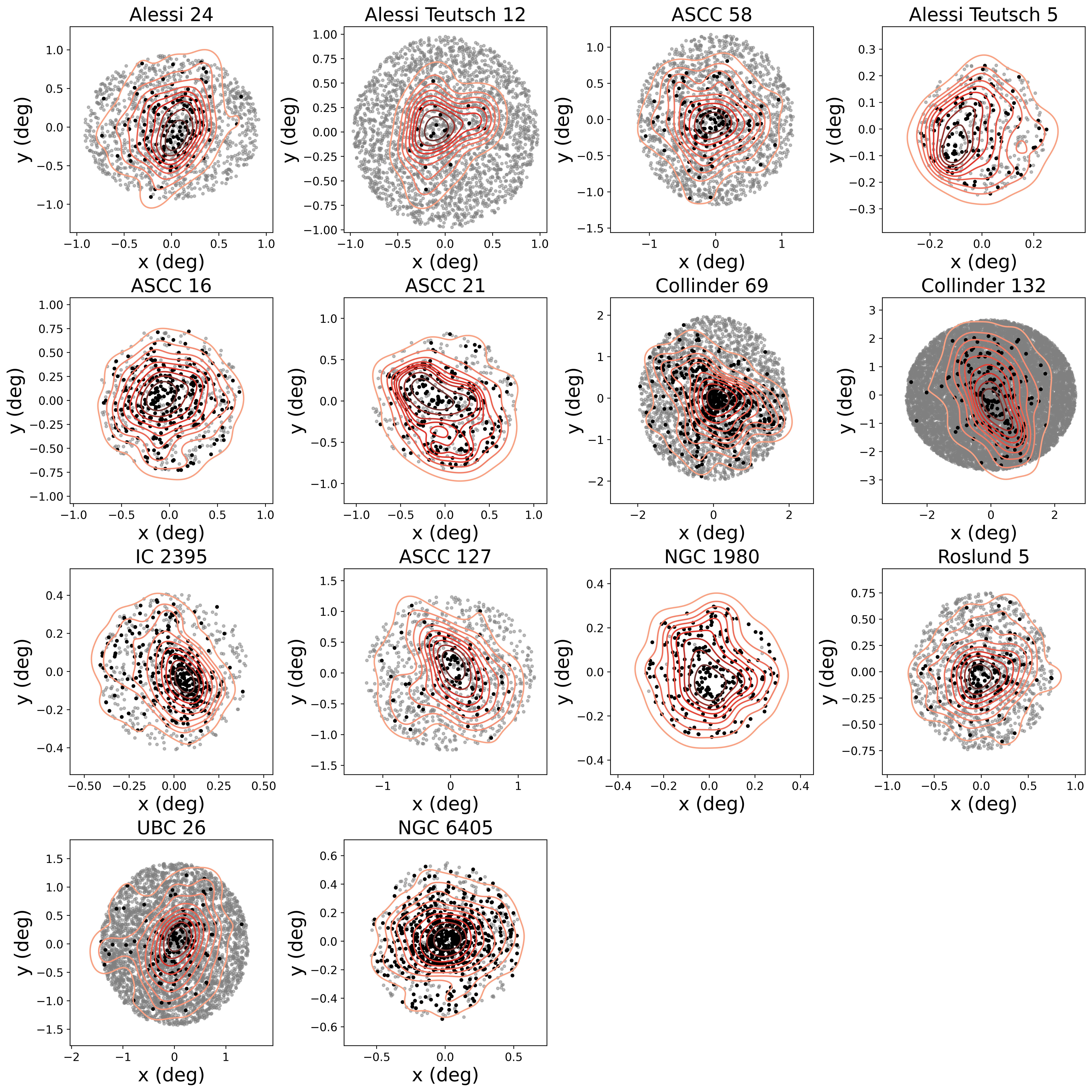}
    \caption{\textit{Spatial plot for all the OCs. Black and gray dots represent the cluster members and field stars, respectively. The red line represents the isodensity contours.}}
    \label{fig:radec}
\end{figure*}

We have determined members of stars belonging to the 14 clusters using GMM as mentioned in the section \ref{sec:GMM}. The number of stars belonging to each cluster is mentioned in Table \ref{table:GMM} and the cluster parameters obtained for each cluster using GMM are shown in Table \ref{table:GMM}.

Spatial plots were plotted to study the spatial distribution of each cluster, which is shown in figure \ref{fig:radec}.  We note that the clusters are located in regions with a wide range of field contamination. Vector Point Diagrams (VPDs)  show the PM distribution of cluster members along with the field stars (Figure \ref{fig:VPD}). From the VPDs, we can see that all the cluster members are identified as a compact group with cluster members having similar PMs. The PM of field stars shows a range. It is interesting to note the paucity of field stars near the cluster NGC 1980. On the other hand, we note a large population of field stars in the field of Collinder 132 and UBC 26.

The number of cluster members estimated from our study is lower than that estimated by \cite{CG} except for 7 OCs (ASCC 16, ASCC 21, Collinder 132, IC 2395, NGC 1980, UBC 26, NGC 6405), Table \ref{Table:Compare}. This difference in numbers can be due to the inclusion of fainter sources between G mag = 18 and 19 in our study, while, \cite{CG} has only considered stars brighter than G mag = 18. A comparison of the number of stars brighter than G mag = 18 with the literature study is shown in figure \ref{fig:lit_compare}. NGC 6405 was studied by \cite{Gao2018} using a combined method of GMM and random forest to compute membership probabilities using \textit{Gaia} GR2, and identified 581 members. In this study, 641 members were identified using \textit{Gaia} DR3 data.

\begin{figure*}
    \centering
    \includegraphics[width=0.95\textwidth]{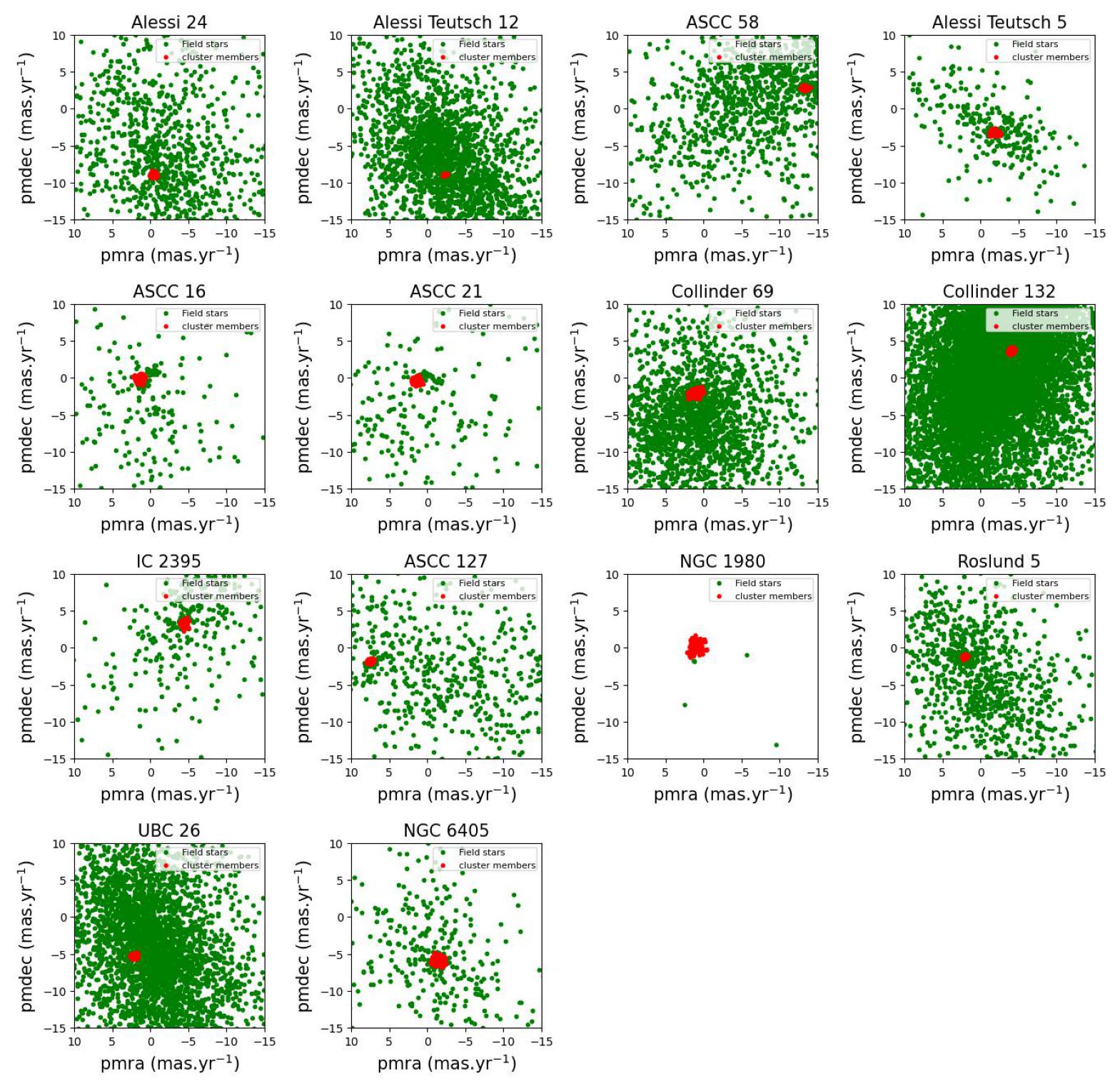}
    \caption{\textit{Vector Point Diagram (VPD) for all the OCs. Red and green dots represent the cluster members and field stars, respectively.}}
    \label{fig:VPD}
\end{figure*}

\subsection{Cluster parameters}
CMDs are used to estimate the cluster parameters such as age, distance, metallicity, and extinction. PARSEC isochrones were fitted against the CMD for each cluster for this estimation, section \ref{sec:iso_fit}. The parameters estimated by least-square minimization and MCMC are tabulated in Table \ref{table_isofit}. The estimated parameters are closer to the values estimated by \cite{CG}. We have also overplotted the CMD and best-fit isochrone to visualize the credibility of the isochrone-fitting method we used. And the isochrone fitted images for least-square minimization and the MCMC methods are shown in figure \ref{fig:iso_fit_chi2} and \ref{fig:iso_fit_mcmc}, respectively.

\begin{figure*}
    \centering
    \includegraphics[width=0.95\linewidth]{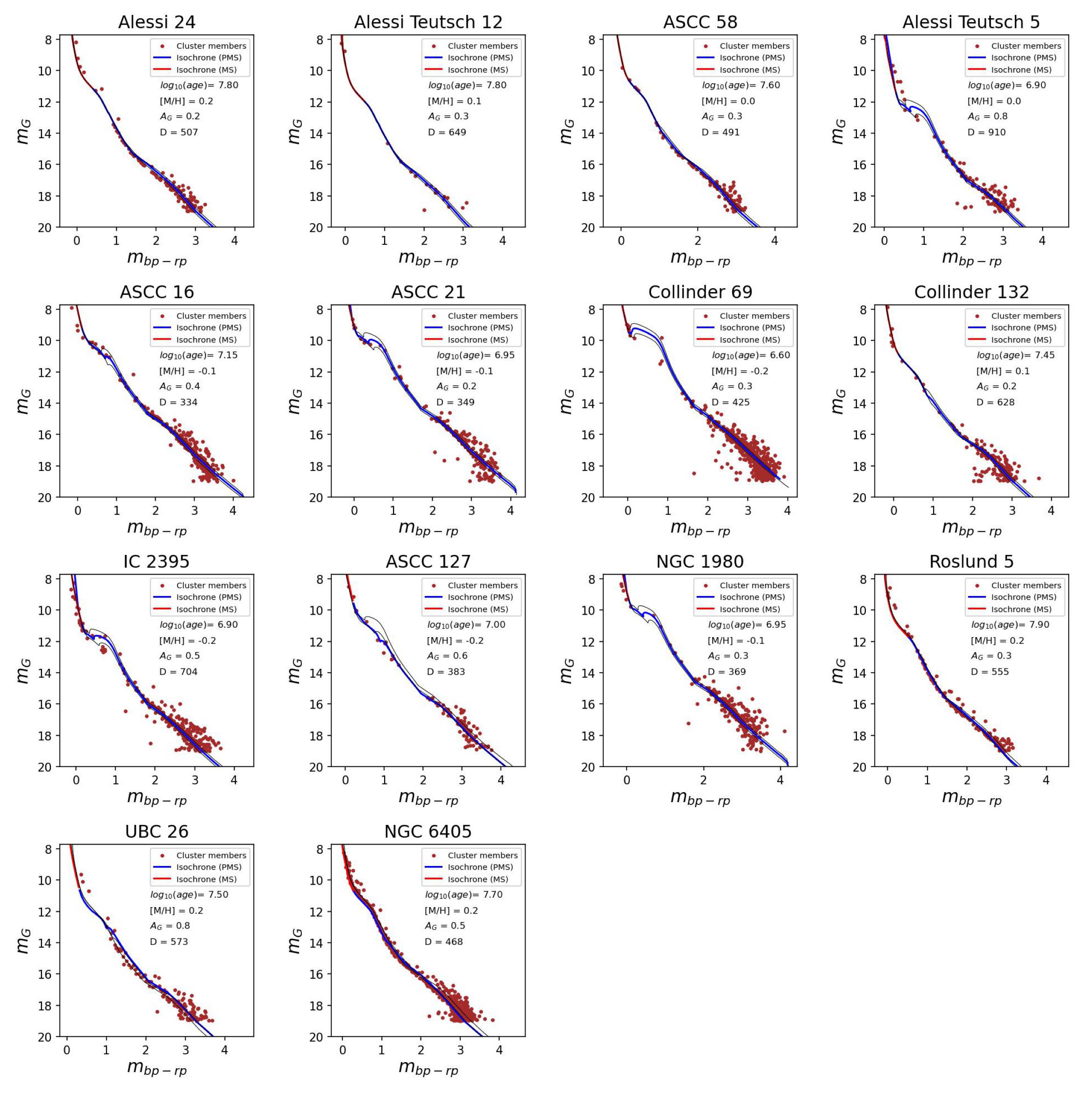}
    \caption{\textit{Isochrone-fitting using least-square minimization for all the OCs. Here, brown dots represent the cluster members; the blue line represents the isochrone for pre-main sequence (PMS) stars, and the red line represents the isochrone for main sequence (MS) stars. Also, $log_{10} (Age)$, Metallicity (M/H), Distance (D) in PC, and Extinction in Gmag ($A_G$) estimated from isochrone fitting are mentioned in the plots. The black lines represent isochrone with $log_{10} (Age) \pm$0.1.}}
    \label{fig:iso_fit_chi2}
\end{figure*}

We used the evolutionary information available in the Isochrone data to classify the cluster members into pre-main sequence (PMS) and Main sequence (MS) stars.
We could see that the majority of stars within each cluster are PMS stars, and very few members are in the MS phase, figure \ref{fig:iso_fit_chi2}, confirming these are all young clusters. The age of the clusters ranges from 7 Myr (Alessi Teutsch 5) to 89 Myr (Roslund 5). Out of the 14 OCs studied, Alessi Teutsch 5 ($910.018^{+0.039}_{-0.024}$ pc) is the farthest, while ASCC 16 ($334.002^{+0.010}_{-0.009} $ pc) is the closest one, Table \ref{table_isofit}. The estimated error of the parameters is lower as we used fiducial points to fit MCMC, assuming that there is no spread in the data set.

The cluster parameters estimated for the OCs, as shown in Table \ref{Table:Compare}, are compared to previous studies, figure \ref{fig:lit_compare}. The distance and age estimates are comparable to the values of \cite{CG, Dias2021}. A higher difference in the parameters estimated compared to \cite{Kharchenko2016} can be due to the difference in the data set used and probable contamination from field stars. With the increased precision of \textit{Gaia} DR3 data, the parameters derived in this study can be considered as better estimates for these clusters.

\subsection{Velocity}
Transverse velocity and its dispersion were then determined using the best-fit parameter obtained for distance from isochrone fitting by using the equation mentioned in section \ref{sec:vel}. The result for this is tabulated in Table \ref{table_sp_vel}. From the analysis, we found that the dispersion in transverse velocity for the Alessi Teutsch 5 cluster is greater than 1 km s$^{-1}$. All the other clusters have transverse velocity dispersion in the range 0.28 -- 0.70 km s$^{-1}$. Literature estimations of transverse velocity dispersion of OCs suggest that they are generally $\le$ 2 km s$^{-1}$ \citep{McNamara1986, RamSagar1989}. Recently, \cite{Kulesh} estimated the velocity dispersion along the Right Ascension and Declination for NGC 2571 to be $\sim 0.55$ km s$^{-1}$. We note that the transverse velocity dispersion for most of the clusters is in this range. 

We also estimated the space velocity of each cluster by using the radial velocity information available in \textit{Gaia} DR3 as mentioned in sec \ref{sec:vel}, Table \ref{table_sp_vel}. Since measurement error overpowers the standard error of radial velocity, we have used measurement error for the error determination of space velocity. Also, due to the limited amount of information on radial velocity available, the calculated space velocity can have a higher error than the calculated one. From the analysis, ASCC 16 has the lowest space velocity of 12.96 $\pm 0.04$ km s$^{-1}$ , and ASCC 58 has the highest space velocity of 34.59 $\pm 0.01$ km s$^{-1}$.

\begin{table}[h]
    \centering
    \renewcommand{\arraystretch}{1.3}
    \resizebox{8.5cm}{!}{
        \begin{tabular}{l l l l}
            \hline
            Cluster name & pmra & pmdec & $N_{star}$\\
            & (mas yr$^{-1}$) & (mas yr$^{-1}$) & \\
            \hline 
            Alessi 24 & $-0.441\pm0.012$& $-8.95\pm0.013$& 135\\
            Alessi Teutsch 12 & $-2.324\pm0.046$& $-8.907\pm0.045$& 22\\
            ASCC 58 & $-13.262\pm0.02$& $2.793\pm0.02$& 123\\
            Alessi Teutsch 5 & $-1.868\pm0.031$& $-3.317\pm0.027$& 118\\
            ASCC 16 & $1.348\pm0.017$& $-0.097\pm0.017$& 280\\
            ASCC 21 & $1.458\pm0.018$& $-0.491\pm0.018$& 223\\
            Collinder 69 & $1.216\pm0.018$& $-2.134\pm0.017$& 517\\
            Collinder 132 & $-4.006\pm0.033$& $3.698\pm0.029$& 186\\
            IC 2395 & $-4.426\pm0.011$& $3.307\pm0.011$& 329\\
            ASCC 127 & $7.471\pm0.02$& $-1.813\pm0.021$& 113\\
            NGC 1980 & $1.173\pm0.031$& $0.374\pm0.031$& 249\\
            Roslund 5 & $2.056\pm0.013$& $-1.212\pm0.013$& 151\\
            UBC 26 & $2.09\pm0.018$& $-5.272\pm0.018$& 140\\
            NGC 6405 & $-1.389\pm0.013$& $-5.874\pm0.011$& 641\\
            \hline
        
        \end{tabular}
    }
    \caption{\textit{Cluster parameters obtained for each cluster using GMM. Columns 2-3 are the two components of PM along with their errors. $N_{star}$ represents the number of sources with probability $\ge$90$\%$}}. 
    \label{table:GMM}
\end{table}

\begin{table*}[h]
    \centering
    \renewcommand{\arraystretch}{1.5}
    \resizebox{17cm}{!}{
        \begin{tabular}{l |l l l l |l l l l}
             \hline
            Cluster name& \multicolumn{4}{c|}{Least-square minimization} & \multicolumn{4}{c}{MCMC}\\
            & D & $log_{10} (age)$ & [M/H] & $A_G$  & D &  $log_{10}(age)$ &[M/H] &$A_G$ \\
             & (pc) & (dex) & (dex) & (mag)  & (pc) &  (dex) &(dex) &(mag) \\
            \hline
            
            Alessi 24         & 507 & 7.8 & 0.2 & 0.2  & $500.160^{+0.970}_{-0.811}$  &  $7.795^{+0.001}_{-0.002}$ &$0.211^{+0.002}_{-0.001}$  &$0.135^{+0.005}_{-0.005}$ \\
            Alessi Teutsch 12 & 649 & 7.8 & 0.1 & 0.3  & $646.998^{+0.015}_{-0.020}$  &  $7.788^{+0.004}_{-0.007}$ &$0.092^{+0.013}_{-0.013}$  &$0.201^{+0.022}_{-0.009}$ \\
            ASCC 58           & 491 & 7.6 & 0 & 0.3  & $490.998^{+0.100}_{-0.185}$  &  $7.601^{+0.005}_{-0.006}$ &$-0.003^{+0.008}_{-0.003}$ &$0.285^{+0.017}_{-0.013}$ \\
            Alessi Teutsch 5  & 910 & 6.9 & 0 & 0.8  & $910.018^{+0.039}_{-0.024}$  &  $6.886^{+0.001}_{-0.016}$ &$0.059^{+0.001}_{-0.036}$  &$0.725^{+0.017}_{-0.011}$ \\
            ASCC 16           & 334 & 7.15 & -0.1 & 0.4  & $334.002^{+0.010}_{-0.009}$  &  $7.157^{+0.005}_{-0.020}$ &$-0.104^{+0.010}_{-0.007}$ &$0.403^{+0.019}_{-0.011}$ \\
            ASCC 21           & 349 & 6.95 & -0.1 & 0.2  & $350.006^{+0.010}_{-0.013}$  &  $7.021^{+0.011}_{-0.001}$ &$-0.107^{+0.016}_{-0.001}$ &$0.193^{+0.006}_{-0.018}$ \\
            Collinder 69      & 425 & 6.6 & -0.2 & 0.3  & $418.243^{+0.047}_{-0.069}$  &  $7.011^{+0.016}_{-0.057}$ &$0.172^{+0.062}_{-0.071}$  &$0.335^{+0.014}_{-0.009}$ \\
            Collinder 132     & 628 & 7.45 & 0.1 & 0.2  & $628.000^{+0.031}_{-0.040}$  &  $7.445^{+0.031}_{-0.013}$ &$0.092^{+0.032}_{-0.004}$  &$0.204^{+0.008}_{-0.023}$ \\
            IC 2395           & 704 & 6.9 & -0.2 & 0.5  & $704.001^{+0.006}_{-0.008}$  &  $6.897^{+0.005}_{-0.001}$ &$-0.204^{+0.009}_{-0.000}$ &$0.504^{+0.002}_{-0.002}$ \\
            ASCC 127          & 383 & 7 & -0.2 & 0.6  & $350.944^{+13.498}_{-3.142}$ &  $7.147^{+0.001}_{-0.001}$ &$-0.254^{+0.004}_{-0.001}$ &$0.724^{+0.019}_{-0.053}$ \\
            NGC 1980          & 369 & 6.95 & -0.1 & 0.3  & $351.984^{+6.310}_{-4.238}$  &  $6.940^{+0.008}_{-0.008}$ &$-0.159^{+0.009}_{-0.005}$ &$0.382^{+0.076}_{-0.078}$ \\
            Roslund 5         & 555 & 7.9 & 0.2 & 0.3  & $555.011^{+0.009}_{-0.017}$  &  $7.946^{+0.008}_{-0.003}$ &$0.194^{+0.016}_{-0.022}$  &$0.309^{+0.010}_{-0.018}$ \\
            UBC 26            & 573 & 7.5 & 0.2 & 0.8  & $578.684^{+0.020}_{-0.012}$  &  $7.482^{+0.023}_{-0.028}$ &$0.211^{+0.034}_{-0.028}$  &$0.771^{+0.025}_{-0.016}$ \\
            NGC 6405          & 468 & 7.7 & 0.2 & 0.5  & $468.003^{+0.020}_{-0.022}$  &  $7.697^{+0.009}_{-0.002}$ &$0.193^{+0.018}_{-0.005}$  &$0.469^{+0.004}_{-0.005}$ \\
            \hline

        \end{tabular}
    }
    \caption{\textit{Isochrone fitting results from least-square minimization and MCMC. D, $log_{10}(age)$, [M/H], and $A_G$ are the distance, $log_{10}(age)$, metallicity, and extinction in G mag, respectively, for least-square minimization and MCMC.}}
    \label{table_isofit}
\end{table*}

\begin{figure*}
   \centering
   \includegraphics[width=1\linewidth]{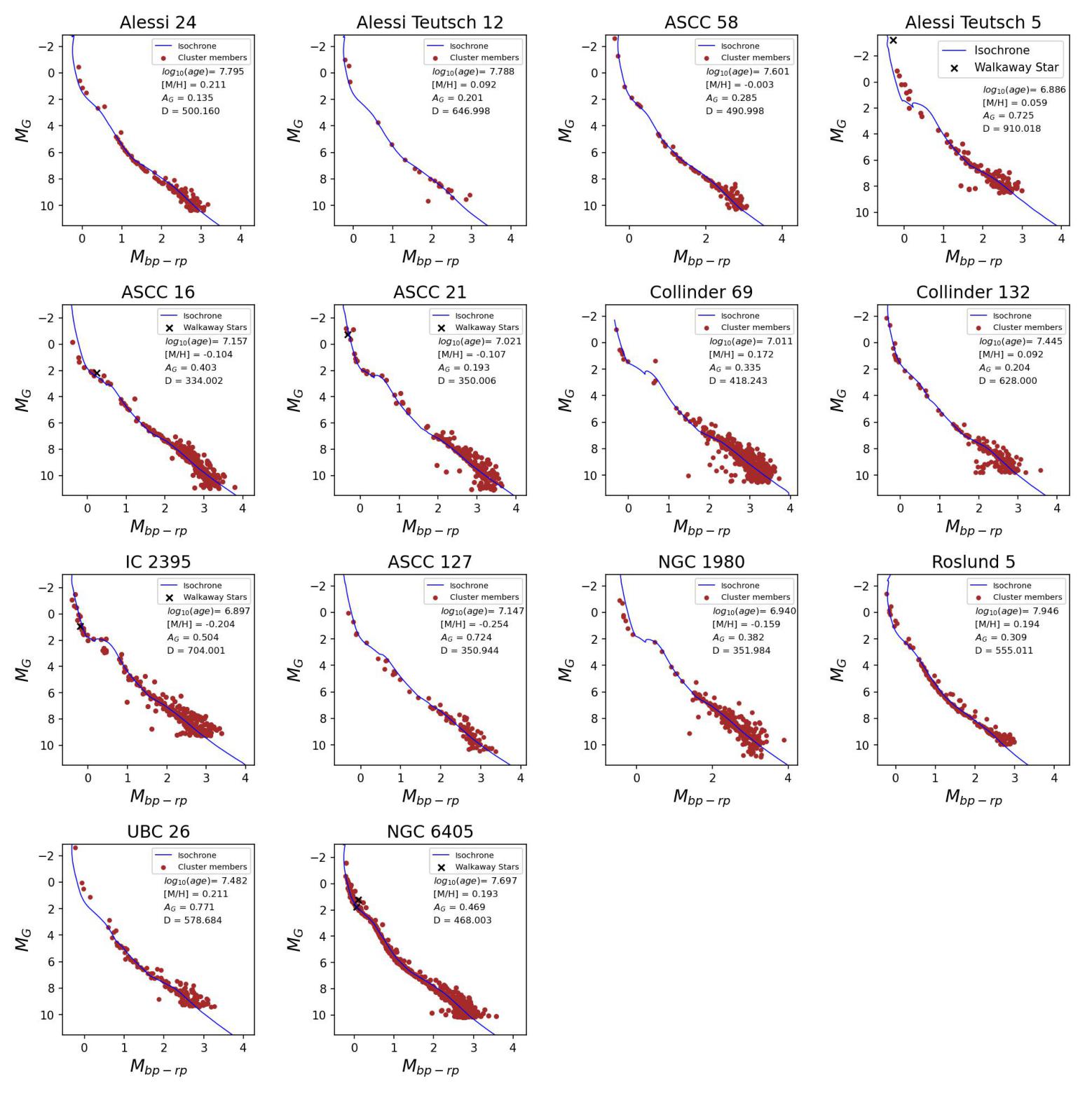}
   \caption{\textit{Isochrone-fitting using MCMC for all the OCs. Here, brown dots represent the cluster members; the blue line represents the isochrone, and the black cross represents the possible walkaway stars.}}
   \label{fig:iso_fit_mcmc}
\end{figure*}

\begin{table*}[h]
    \centering
    \renewcommand{\arraystretch}{1.3}
    \resizebox{16cm}{!}{
        \begin{tabular}{l l l l l l l l l l l l}
            \hline
            Cluster name & $N_{RV}$ & RV\_mean & RV\_disp & RV\_std\_er & RV\_er & $N_{star}$ & Tvel\_mean & Tvel\_disp & Tvel\_std\_er & Space\_V & Space\_V\_er \\
            & & (km s$^{-1}$) & (km s$^{-1}$) & (km s$^{-1}$) & (km s$^{-1}$) & & (km s$^{-1}$) & (km s$^{-1}$) & (km s$^{-1}$) & (km s$^{-1}$) & (km s$^{-1}$)\\
            \hline
            Alessi 24 & 17& 10.20& 4.89& 1.19& 4.12& 135& 21.25& 0.43& 0.04& 23.57& 0.02\\
            Alessi Teutsch 12 & 3& -3.95& 3.33& 1.92& 4.86& 22& 28.26& 0.28& 0.06& 28.54& 0.04\\
            ASCC 58 & 14& 14.15& 11.34& 3.03& 4.74& 123& 31.57& 0.47& 0.04& 34.59& 0.01\\
            Alessi Teutsch 5 & 6& 0.27& 28.41& 11.60& 11.45& 118& 16.68& 1.22& 0.11& 16.68& 2.53\\
            ASCC 16 & 55& 12.77& 16.26& 2.19& 5.93& 280& 2.20& 0.43& 0.03& 12.96& 0.04\\
            ASCC 21 & 37& 13.74& 25.49& 4.19& 6.01& 223& 2.59& 0.38& 0.03& 13.98& 0.03\\
            Collinder 69 & 37& 17.85& 22.59& 3.71& 6.85& 517& 4.88& 0.67& 0.03& 18.51& 0.02\\
            Collinder 132 & 17& 21.75& 5.40& 1.31& 5.00& 186& 16.50& 0.65& 0.05& 27.30& 0.01\\
            IC 2395 & 35& 24.34& 17.76& 3.00& 8.06& 329& 18.39& 0.70& 0.04& 30.50& 0.01\\
            ASCC 127 & 16& -13.86& 16.93& 4.23& 4.40& 113& 12.81& 0.33& 0.03& 18.87& 0.02\\
            NGC 1980 & 36& 22.30& 24.54& 4.09& 6.21& 249& 2.28& 0.56& 0.04& 22.42& 0.01\\
            Roslund 5 & 38& -17.53& 5.51& 0.89& 4.05& 151& 6.32& 0.41& 0.03& 18.63& 0.01\\
            UBC 26 & 14& -15.12& 13.34& 3.56& 4.68& 140& 15.65& 0.57& 0.05& 21.76& 0.01\\
            NGC 6405 & 164& -11.06& 8.81& 0.69& 4.60& 641& 13.43& 0.69& 0.03& 17.40& 0.02\\
            \hline
        \end{tabular}
    }
    \caption{\textit{Radial velocity and space velocity estimation for the clusters. $N_{RV}$ represents the number of stars in a cluster with radial velocity information available in the \textit{Gaia} DR3. Columns 3--6 represent the Radial velocity mean and its dispersion, standard error, and measurement error, respectively. $N_{star}$ is the number of cluster members. Columns 8--10 are the Transverse velocity mean and its dispersion and standard error, respectively. Columns 11 and 12 represent the space velocity and its error.}}
    \label{table_sp_vel}

\end{table*}

\begin{table*}[h]
    \centering
    \renewcommand{\arraystretch}{1.3}
    \resizebox{15cm}{!}{
        \begin{tabular}{l l l l l l l l l}
            \hline
            Cluster name     & Mass\_BA & Mass\_FG & Mass\_K & Mass\_M & BA\_Tvel\_disp & FG\_Tvel\_disp & K\_Tvel\_disp & M\_Tvel\_disp \\
            & ($M_\odot$) & ($M_\odot$) & ($M_\odot$) & ($M_\odot$) & (km s$^{-1}$) & (km s$^{-1}$) & (km s$^{-1}$) & (km s$^{-1}$) \\
            \hline
            Alessi 24         & 3.6& 1.21& 0.74& 0.47& $0.21\pm0.09$& $0.21\pm0.08$& $0.29\pm0.06$& $0.47\pm0.05$\\
            Alessi Teutsch 12 & 4.55& 1.11& 0.68& 0.53& $0.19\pm0.09$& $0.29\pm0.2$& $0.19\pm0.09$& $0.31\pm0.09$\\
            ASCC 58           & 3.45& 1.11& 0.72& 0.43& $0.38\pm0.15$& $0.18\pm0.08$& $0.31\pm0.07$& $0.5\pm0.05$\\
            Alessi Teutsch 5  & 4.24& 1.26& 0.65& 0.45& $1.44\pm0.46$& $1.06\pm0.3$& $1.03\pm0.12$& $1.67\pm0.36$\\
            ASCC 16           & 3.04& 1.29& 0.68& 0.31& $0.4\pm0.12$& $0.29\pm0.08$& $0.34\pm0.04$& $0.46\pm0.03$\\
            ASCC 21           & 3.99& 1.23& 0.64& 0.29& $0.35\pm0.09$& $0.28\pm0.1$& $0.32\pm0.05$& $0.4\pm0.03$\\
            Collinder 69      & 3.11& 1.31& 0.7& 0.34& $0.44\pm0.15$& $0.49\pm0.22$& $0.6\pm0.05$& $0.69\pm0.04$\\
            Collinder 132     & 3.32& 1.27& 0.73& 0.47& $0.26\pm0.08$& $0.27\pm0.09$& $0.56\pm0.09$& $0.69\pm0.06$\\
            IC 2395           & 3.17& 1.26& 0.58& 0.32& $0.78\pm0.17$& $0.47\pm0.09$& $0.61\pm0.05$& $0.79\pm0.06$\\
            ASCC 127          & 2.59& 1.11& 0.62& 0.29& $0.14\pm0.06$& $0.26\pm0.09$& $0.31\pm0.06$& $0.35\pm0.04$\\
            NGC 1980          & 4.62& 1.21& 0.6& 0.28& $0.54\pm0.18$& $0.27\pm0.11$& $0.52\pm0.05$& $0.59\pm0.05$\\
            Roslund 5         & 3.08& 1.17& 0.74& 0.49& $0.31\pm0.1$& $0.35\pm0.07$& $0.35\pm0.05$& $0.49\pm0.06$\\
            UBC 26            & 4.25& 1.13& 0.76& 0.53& $0.49\pm0.25$& $0.56\pm0.15$& $0.46\pm0.07$& $0.6\pm0.07$\\
            NGC 6405          & 2.63& 1.23& 0.76& 0.47& $0.67\pm0.09$& $0.62\pm0.07$& $0.61\pm0.06$& $0.72\pm0.04$\\
            \hline

        \end{tabular}
    }
    \caption{\textit{Transverse velocity and its dispersion for each spectral type. Columns 2--5 represent the mean mass of each spectral type, and columns 6--9 represent the dispersion in transverse velocity along with their errors for the mentioned group of spectral types.}}
    \label{table_obfdisp}
\end{table*}

\begin{table*}[h]
    \centering
    \renewcommand{\arraystretch}{1.3}
    \resizebox{14cm}{!}{
        \begin{tabular}{l l l l l l l l l lll}
            \hline
            Cluster name     & B & A & F & G & K & M & Mass\_min & Mass\_max  & M/K& M/K Salpeter&M/K Kroupa\\
             &  &  &  &  &  &  & ($M_\odot$) & ($M_\odot$)  &  & &\\
            \hline
            Alessi 24         & 3& 2& 2& 5& 26& 97& 0.33& 6.23&$\ge$3.73&2.84&1.32\\
            Alessi Teutsch 12 & 4& 0& 1& 1& 4& 12& 0.43& 6.28&$\ge$3.00&1.77&1.75\\
            ASCC 58           & 2& 4& 1& 4& 23& 89& 0.28& 7.19&$\ge$3.87&3.41&1.42\\
            Alessi Teutsch 5  & 7& 3& 3& 9& 75& 21& 0.41& 13.01&$\ge$0.28&2.37&1.05\\
            ASCC 16           & 2& 9& 8& 5& 63& 193& 0.15& 11.71&$\ge$3.06&6.28&1.89\\
            ASCC 21           & 7& 8& 3& 5& 47& 153& 0.13& 12.89&$\ge$3.26&6.62&1.81\\
            Collinder 69      & 5& 4& 2& 3& 127& 376& 0.20& 5.70&$\ge$2.96&5.31&1.75\\
            Collinder 132     & 4& 7& 6& 2& 38& 129& 0.33& 6.87&$\ge$3.39&2.87&1.31\\
            IC 2395           & 8& 13& 12& 14& 135& 147& 0.25& 7.41&$\ge$1.09&4.06&1.27\\
            ASCC 127          & 2& 3& 3& 5& 24& 76& 0.16& 3.93&$\ge$3.17&6.01&1.67\\
            NGC 1980          & 6& 3& 2& 4& 97& 137& 0.13& 13.77&$\ge$1.41&5.73&1.57\\
            Roslund 5         & 7& 3& 10& 17& 51& 63& 0.41& 4.77&$\ge$1.24&2.63&1.27\\
            UBC 26            & 3& 1& 5& 10& 41& 80& 0.42& 7.70&$\ge$1.95&2.33&2.29\\
            NGC 6405          & 19& 43& 42& 43& 119& 375& 0.34& 7.12&$\ge$3.15&3.04&1.41\\
            \hline
        \end{tabular}
    }
    \caption{\textit{Spectral type classification of cluster members. Columns 2--7 represent the number of sources belonging to each spectral type for the corresponding clusters mentioned in column 1. Columns 8 and 9 represent the mass range of the cluster, with  Mass\_min representing minimum mass and Mass\_max representing maximum mass in solar mass ($M_\odot$). Column 10 represents the observed ratio of M-type stars to the K-type stars in corresponding clusters, which  are the lower bound values. Columns 11 and 12 represent expected M/K ratios from Salpeter \citep{Salpeter1955} and Kroupa \citep{Kroupa2001}, respectively. }}
    \label{table_obf}
\end{table*}

\begin{figure*}
    \centering
    \subfloat[] {\includegraphics[width=0.5\linewidth]{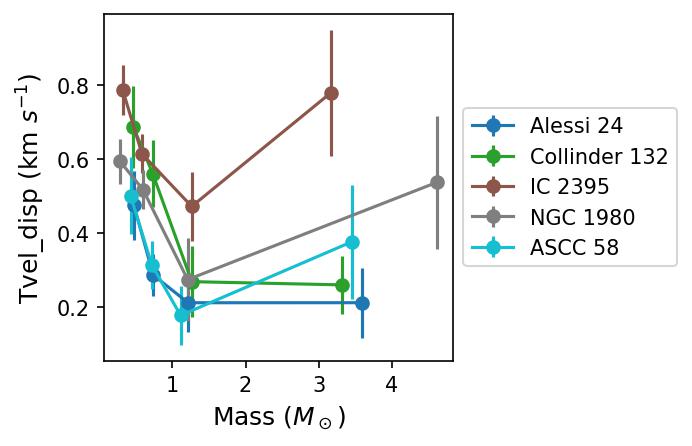}}
    \subfloat[]{\includegraphics[width=0.5\linewidth]{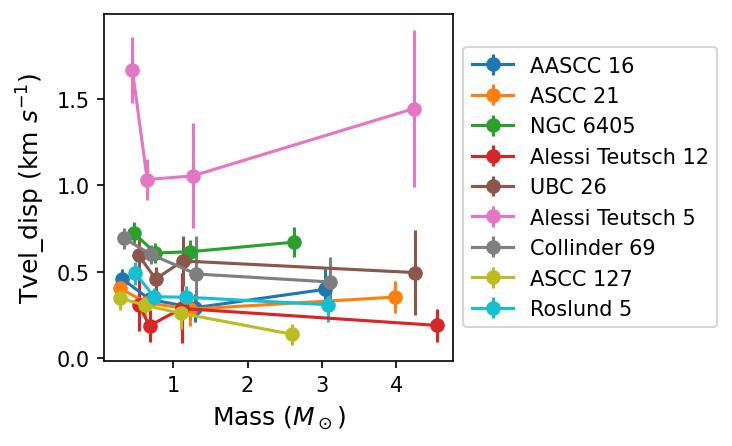}}
    \caption{\textit{Mass vs Transverse velocity dispersion. (a) Statistically significant ones (b) May have statistical insignificance.}}
    \label{fig:mass_vs_disp}
\end{figure*}

\subsection{ Mass dependency of velocity dispersion}
We classified cluster members based on their spectral type (O, B, A, F, G, K, and M) by using the updated version of Erik Mamajek's table, \cite{mamajek}. This is achieved by comparing G mag range for each spectral type in the table to the G mag values our cluster members. They are grouped into BA, FG, K, and M-type stars since the number of stars in the high massive stars (B, A, F, G) is much lower compared to low massive stars (K and M). The mass of the group is obtained by taking the mean mass of stars within a group, and the obtained results are tabulated in Table \ref{table_obfdisp}. The transverse velocity dispersion of each group is determined by taking the standard deviation of transverse velocity within each group and its error as standard error, Table \ref{table_obfdisp}.

Within the mass range of 1.0 to 0.5 M$_\odot$, 5 clusters show a statistically significant increase in the transverse velocity dispersion, figure \ref{fig:mass_vs_disp}. We consider these as statistically significant increases as their error range does not overlap with the mean value of other groups within 2 sigma. These are Collinder 132, NGC 1980, IC 2395, ASCC 58, and Alessi 24. The other clusters either show no trend or the increase is not statistically significant. The increasing velocity dispersion is indicative of kinematic relaxation with the presence of equipartition activity in the cluster. The clusters have a large range in age (8 -- 63 Myr) and have a range in the number of member stars  (123 -- 329 members). The evidence of relaxation for low-mass stars in young clusters such as IC 2395 (8 Myr) and NGC 1980 (9 Myr) are interesting. A similar work was carried out by \cite{RamSagar1989} and \cite{McNamara1986} to study the internal kinematics of the OCs. A Statistically significant increase in transverse velocity dispersion as mass decreases also points to the mass segregation within OCs. The Alessi 24 OC was studied by \cite{Bhattacharya-Kaushar2022}, where they indicated the possibility of slight mass segregation, which agrees with our findings based on velocity dispersion.

\begin{figure*}
    \centering
    \includegraphics[width=0.95\linewidth]{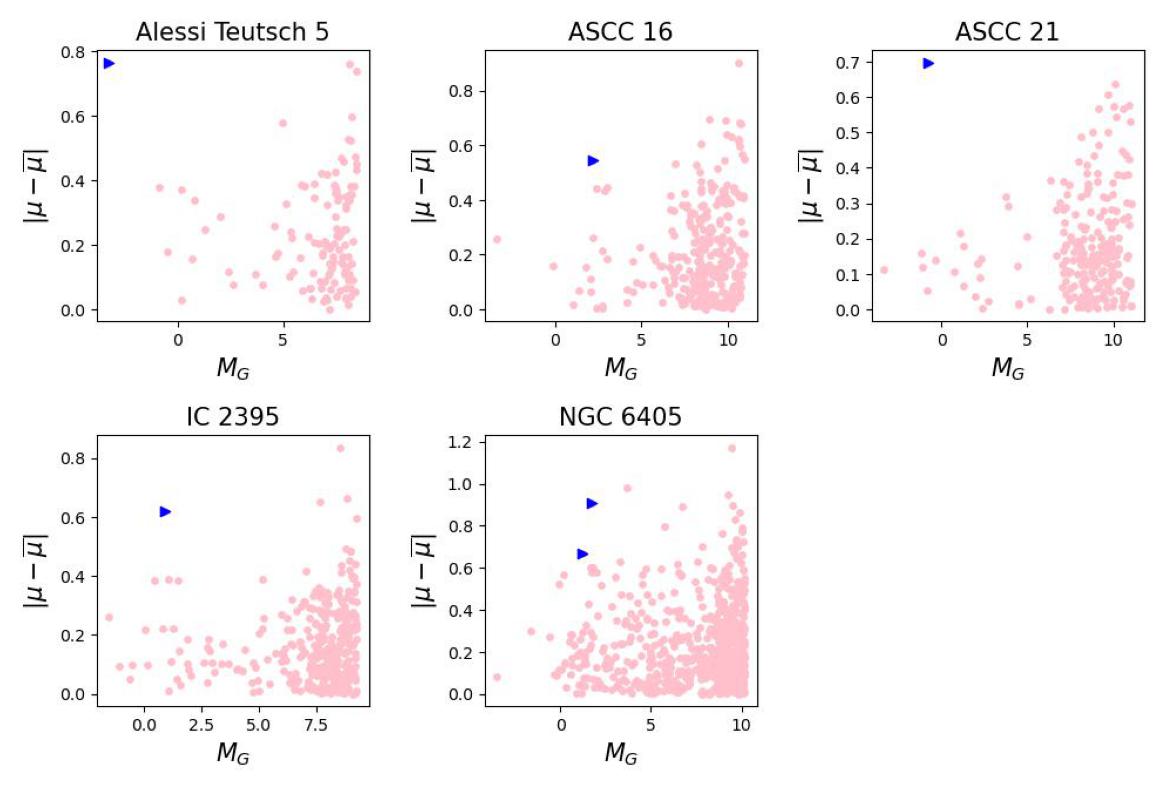}
    \caption{\textit{Clusters with higher massive stars having relatively large variations in their PM. Blue triangle representing the star with higher variation in the PM}}
    \label{fig:walkaway}
\end{figure*}

\begin{figure*}[t]
    \centering
    \includegraphics[width=0.95\linewidth]{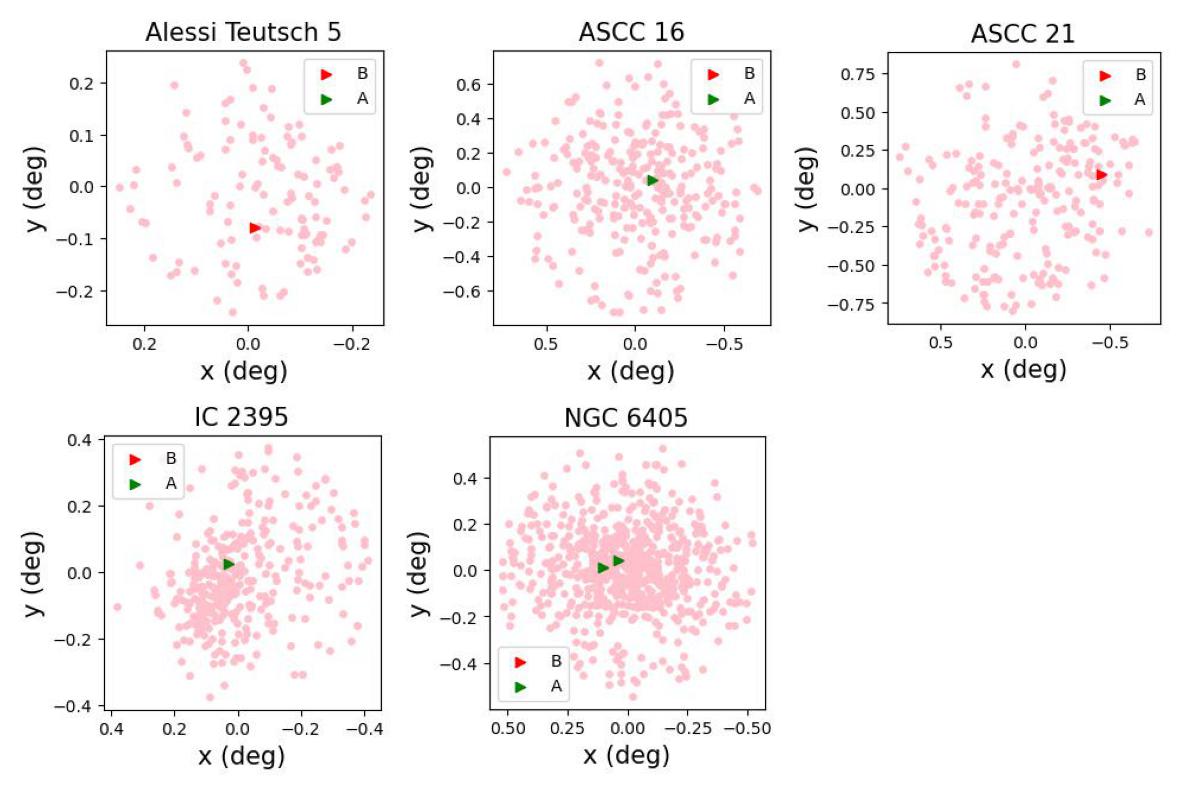}
    \caption{\textit{Spatial plot of clusters with possible walkaway stars detected. The triangle represents the position of stars with higher variation in PM. The red and green triangles represent the B and A-type stars, respectively.}}
    \label{fig:walkaway_spatial}
\end{figure*}

\subsection { High PM massive stars}
In general, low-mass stars show relatively higher velocity dispersion than the high-mass stars. Apart from this, in some clusters, even the high-mass stars appear to show relatively large velocity dispersion. This deviation in high-mass stars not following the velocity-mass relation is in agreement with \cite{RamSagar1989} and \cite{McNamara1986}.

 Some clusters are found to have BA-type stars with relatively large PM, as shown in figure \ref{fig:walkaway}. These are the ones having a transverse velocity greater than 2$\sigma$ from the mean transverse velocity of the OCs. These are found in 5 clusters (Alessi Teutsch 5, ASCC 16, ASCC 21, IC 2395, and NGC 6405), and all these clusters are in the age range 8--15 Myr, except NGC 6405. NGC 6405 (49--50 Myr) is an older cluster, and it has 2 such stars. The relative velocity of these stars ranges from 0.86 -- 3.3 km s$^{-1}$. Also, the mass of these stars ranges between 1.7 -- 13.7 M$_\odot$. The star with the highest relative transverse velocity is seen in Alessi Teutsch 5 (3.3 km s$^{-1}$), which has a mass of 13 M$_\odot$ (B-type star). The locations of these stars are shown in the figure \ref{fig:walkaway_spatial}. Many are found in the central regions of these young clusters. These are potential candidates to escape the cluster potential. \cite{Fujii-Zwart} suggested that the dynamically ejected runaway stars are generally produced in the dense regions of clusters. It is possible that dynamical effects are at work in these 5 clusters responsible for the creation of these large PM massive stars. It is also possible that some high velocity stars have left the cluster already. We suggest that these stars can be possible walkaway stars, which might move out of the cluster in the future. 

\subsection{Fraction of M type to K type stars}
Though we have the membership and mass range of stars, due to incompleteness and poor statistics, it may not be appropriate to estimate the Initial Mass Function (IMF) of these clusters. There are many studies that are dedicated to study the IMF in young clusters such as \cite{Damian2021, Luhman2016, Suarez2019}. On the other hand, we do have a good number of stars in the K and M spectral types, which could be used to check the fraction of stars expected, particularly near the transition mass of $\sim$ 0.4--0.5 M$_\odot$. This can throw light on possible reasons for the variation in the number of stars in the K and M spectral types as seen in the Table \ref{table_obf}. We estimated the present-day M/K ratio for all clusters and compared the value with that expected from (1) Salpeter slope for initial mass function ($-$2.35 for both M \& K type stars, \citeauthor{Salpeter1955} \citeyear{Salpeter1955}) and (2) Kroupa slope for the initial mass function ($-$1.3 for M$_\odot$ $<$ 0.5 and  $-$2.3 for M$_\odot$ $>$ 0.5, \citeauthor{Kroupa2001} \citeyear{Kroupa2001}). The observed M/K values can be the lower bound values due to the incompleteness of the data set. By definition, the expected ratio for Salpeter IMF is higher than that for Kroupa IMF (Figure \ref{fig:mk}). 

Alessi Teutsch 5 has a ratio less than 1, probably because of missing low mass stars due to its large distance. Furthermore, due to the presence of another cluster(BDSB 30), \citep{CG}, closer to Alessi Teutsch 5, we restricted our study to a lower radius, which might also contribute to this lower ratio.

UBC 26 is a nearby young cluster, and the observed and expected ratios are similar. IC 2395, NGC 1980, and Roslund 5 have their ratio similar to the Kroupa ratio (figure \ref{fig:mk}). The M/K ratio estimated for ASCC 16, ASCC 21, ASCC 127, and Collinder 69 appear to be in between that expected from Salpeter and Kroupa IMFs. We expect these ratios might actually be higher than the ones we found as the M/K ratio we found represents the lower bound. The ratios estimated for Alessi 24, ASCC 58, and Collinder 132 appear to be higher than that is expected from even Salpeter IMF, Figure \ref{fig:mk}. These clusters are older, with ages in the range 28 -- 62 Myr. NGC 6405 is an older cluster having a ratio similar to that expected from salpeter IMF. A detailed study needs to be performed to validate this, and we plan to carry this out in the future.

\begin{figure}
    \centering
    \includegraphics[width=1\linewidth]{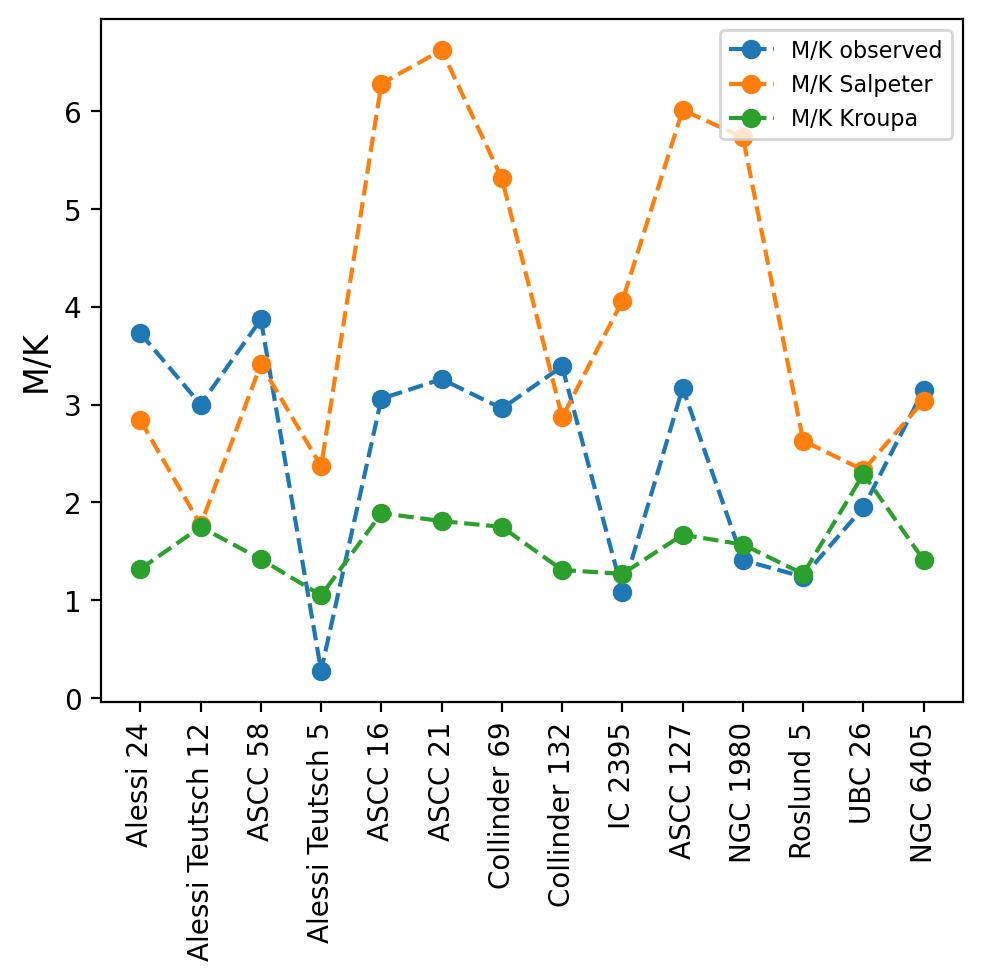}
    \caption{\textit{Comparison of M/K ratio to the ratios of Salpeter and Kroupa}}
    \label{fig:mk}
\end{figure}

\begin{figure*}
    \centering
    \includegraphics[width=0.9\linewidth]{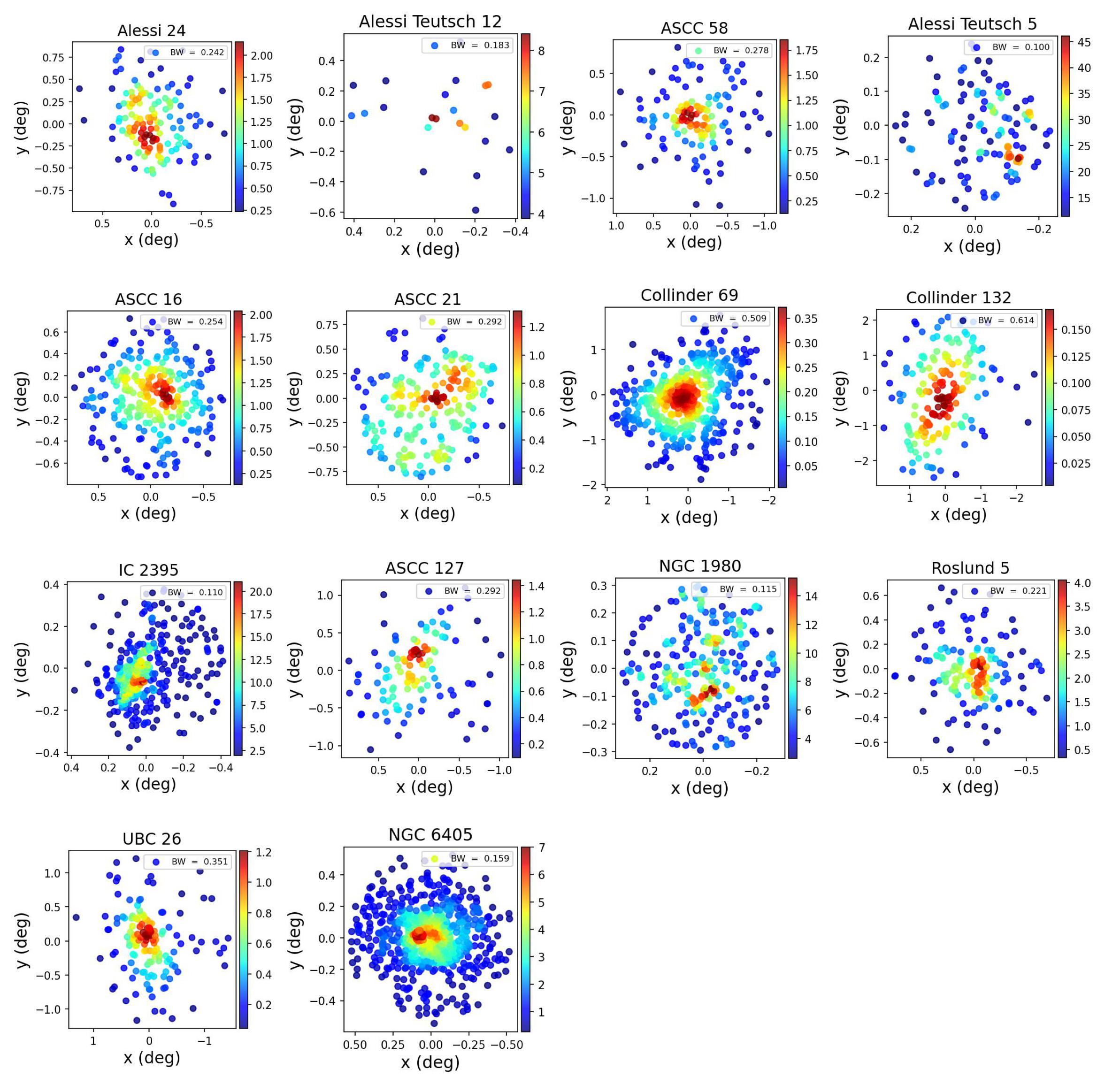}
    \caption{\textit{Kernal density diagrams for each cluster with the Bandwidth (BW) used for plotting mentioned in the top right corner.}}
    \label{fig:kde}
\end{figure*}

\subsection{Structural study of clusters}
An elongated structure is seen for Alessi 24, Collinder 132, ASCC 127, and UBC 26, Figure \ref{fig:kde}, with an age range of 14 to 62 Myr. These clusters are comparatively older (28 -- 62 Myr) except ASCC 127 (14 Myr). This agrees with the suggestion that older clusters become more elliptical compared to younger ones, \cite{Zhai2017}. 
Also, there are two clumps within Alessi 24. The difference in transverse velocity between the clumps is not found to be significant, suggesting that these clumps are not kinematically distinct in the sky-plane. Radial velocity data is needed to estimate whether these clumps will merge or separate in the future. We also note some amount of sub-structures in the full age range of clusters studied here. The spatial sub-structures found in clusters might indicate that clusters retain these for at least up to about 100 Myr.

\subsection{Binary clusters}
The fraction of binary clusters in the Milky Way Galaxy is 8\% to 20\% \citep{Subramaniam1995, Marcos2009, Soubiran2018}. ASCC 16 and ASCC 21 have been previously identified as a binary cluster by \cite{Soubiran2018}. They have a similar age, log(age) = $7.157^{+0.005}_{-0.020}$, $7.021^{+0.011}_{-0.001}$ respectively, and similar metallicity,  [M/H] = $-0.104^{+0.010}_{-0.007}$, $-0.107^{+0.016}_{-0.001}$ respectively. They have a transverse velocity difference of 0.39 km s$^{-1}$ and a space velocity difference of 1.02 km s$^{-1}$. Furthermore, their separation of 43.32 pc further suggests that these two clusters might have originated from the same molecular cloud around the same time. 


\section{Conclusions and Summary} \label{sec:Conclusion}
In this study, we present the kinematic and structural properties of 14 clusters using \textit{Gaia} DR3 to link them to the details of internal dynamics and imprints of star formation in young clusters. The summary and conclusions of this study are given below:

\begin{itemize}
    \item GMM method was used to find the membership information of the OCs. The clusters chosen are young ($<$ 100 Myr) and within a distance of 1 kpc, in order to study the internal kinematics as a function of mass within the clusters. The cluster parameters, such as age, distance, metallicity, and extinction, were estimated by fitting PARSEC isochrones to the CMDs, using a combination of a least-square minimization and MCMC technique. The code is automated to apply to a larger sample of clusters in the future. 

    \item The clusters are found to be located between 350 - 910 pc, with an age ranging from 7.7 Myr - 88 Myr, and near solar metallicity. The number of member stars ranges from 22 to 641. Most of the low-mass (K and M-type) stars in the cluster are in the pre-main sequence (PMS) phase, whereas the high-mass stars are found to be in the main-sequence phase.
    
    \item 13 clusters are found to have a transverse velocity dispersion in the range 0.28 - 0.70 km s$^{-1}$, with 10 clusters between 0.41 - 0.70 km s$^{-1}$. We suggest that this is the typical transverse velocity dispersion range in young clusters. 
    
    \item Alessi Teutsch 12 has only 22 members, but has a low transverse velocity dispersion of $\sim$ 0.28$\pm$0.06 km s$^{-1}$,  suggesting that it is a poor open cluster. The largest transverse velocity dispersion 1.22$\pm$0.11 km s$^{-1}$ is found in Alessi Teutsch 5, with 118 members.
    
    \item The rich clusters in our sample, Collinder 69 (517 members), NGC 6405 (641 members), and IC 2395 (329 members) show a similar transverse velocity dispersion of $\sim$ 0.67 -- 0.70 km s$^{- 1}$. 

    \item All clusters have stars in the B - M spectral type range. Alessi Teutsch 5, ASCC 16, ASCC21, and NGC 1980 have stars more massive than 10 M$_\odot$. NGC 6405 has the largest number of B-type stars (19).

    \item Five clusters (Alessi 24, Collinder 132, IC 2395, NGC 1980, and ASCC 58) show increasing velocity dispersion from F to M-type stars, a signature for dynamical relaxation for low-mass stars. The evidence of relaxation for low-mass stars in young clusters such as IC 2395 (8 Myr) and NGC 1980 (9 Myr) are interesting.

    \item In the above clusters, we found that the B-A spectral types appear to have relatively higher or the same velocity dispersion as the FG type stars, suggesting the velocity-mass relation does not strictly hold. We are unsure about the cause of this phenomenon. It could be partially due to the low number of massive stars and/or due to the dynamical effects of binaries. 
    
    \item This study, therefore, provides probably the first quantitative estimation of mass-dependent velocity dispersion as a pointer to dynamical relaxation in young clusters through transverse velocity dispersion estimates. This is an important pointer for numerical simulations of cluster dynamics, escape of low-mass stars, mass segregation, etc. 

    \item We have found possible walkaway stars in 5 clusters (Alessi Teutsch 5, ASCC 16, ASCC 21, IC 2395, and NGC 6405). The highest relative velocity (3.3 km s$^{-1}$) is found for a 13 M$_\odot$ B-type star in Alessi Teutsch 5. The reason why we are not able to see stars with much higher velocities can possibly be due to the clustering algorithm (GMM) we have used, which by default reduces the posterior probability of the stars with higher PM.

    \item We estimated the ratio of M/K type stars and compared them with the values expected from Kroupa and Salpeter IMF slopes.
    We note that the observed M/K ratio shows a trend with age, which might indicate the missing detection of M-type stars in younger clusters, resulting in a lower value of the ratio. On the other hand, the higher values found in 3 clusters (Alessi 24, ASCC 58, and Collinder 132, even with a higher transverse velocity dispersion for M-type stars) are likely to be real, but need more detailed study.

    \item The Spatial distribution of the members shows patterns, and we note clumpiness in a few clusters, including Alessi 24, which are kinematically indistinguishable. Elongated structures are seen in comparatively older clusters. The spatial sub-structures found in clusters might indicate that clusters retain these structures at least up to $\sim$ 90 Myr.

    \item We find that ASCC 16 ($\sim$ 14 Myr) and ASCC 21 ($\sim$ 10.5 Myr), suggested to be a binary cluster candidate, have a transverse velocity difference of 0.39 km s$^{-1}$ and a space velocity difference of 1.02 km s$^{-1}$. Furthermore, their separation of 43.32 pc further suggests that these two clusters might have formed from the same molecular cloud, but are unlikely to be a binary.
    
\end{itemize}

     
\clearpage
\onecolumn
\appendix

\renewcommand{\thefigure}{A\arabic{figure}}
\setcounter{figure}{0}
\renewcommand{\thetable}{A\arabic{table}}
\setcounter{table}{0}

\section{Literature survey}

\begin{table}[h]
    \centering
    \renewcommand{\arraystretch}{1.3}
    \resizebox{14cm}{!}{
        \begin{tabular}{l l l l l l l l l l}
            \hline
            Cluster name & R.A.$^*$ & Dec.$^*$ & pmra$^*$ & pmdec$^*$ & D$^*$ & r50$^*$ & Rmax & r50$\times$2 & $log_{10}(age)^*$ \\
            & (deg) & (deg) & (mas yr$^{-1}$) & (mas yr$^{-1}$) & (kpc) & (deg) & (arcmin) & (deg) &  \\
            \hline
            Alessi 24 & 260.764 & -62.693 & -0.488 & -8.999 & 0.502 & 0.466 & 55.92 & 0.932 & 7.86 \\
            Alessi Teutsch 12 & 255.421 & -58.981 & -2.134 & -8.952 & 0.641 & 0.489 & 58.68 & 0.978 & 7.94 \\
            ASCC 58 & 153.657 & -55.001 & -13.276 & 2.786 & 0.491 & 0.592 & 71.04 & 1.184 & 7.72 \\
            Alessi Teutsch 5 & 332.218 & 61.103 & -1.895 & -3.207 & 0.907 & 0.125 & 15 & 0.25 & 7.27 \\
            ASCC 16 & 81.198 & 1.655 & 1.355 & -0.015 & 0.344 & 0.376 & 45.12 & 0.752 & 7.13 \\
            ASCC 21 & 82.179 & 3.527 & 1.404 & -0.632 & 0.341 & 0.41 & 49.2 & 0.82 & 6.95 \\
            Collinder 69 & 83.792 & 9.813 & 1.194 & -2.118 & 0.416 & 0.989 & 118.68 & 1.978 & 7.1 \\
            Collinder 132 & 108.485 & -30.758 & -4.14 & 3.732 & 0.624 & 1.33 & 159.6 & 2.66 & 7.39 \\
            IC 2395 & 130.531 & -48.09 & -4.464 & 3.293 & 0.702 & 0.209 & 25.08 & 0.418 & 7.31 \\
            ASCC 127 & 347.205 & +64.974 & 7.474 & -1.745 & 0.376 & 0.627 & 75.24 & 1.254 & 7.26 \\
            NGC 1980 & 83.81 & -5.924 & 1.215 & 0.54 & 0.377 & 0.151 & 18.12 & 0.302 & 7.12 \\
            Roslund 5 & 302.641 & +33.751 & 1.981 & -1.164 & 0.546 & 0.378 & 45.36 & 0.756 & 7.99 \\
            UBC 26 & 285.37 & +22.020 & 2.049 & -5.176 & 0.582 & 0.717 & 86.04 & 1.434 & 7.55 \\
            NGC 6405 & 265.069 & -32.242 & -1.306 & -5.847 & 0.459 & 0.275 & 33 & 0.55 & 7.54 \\
            \hline
        \end{tabular}
     }
    \vspace{-3mm}
    
    \caption*{\small\textit{Note: $^*$ obtained from Cantat-Gaudin catalogue }}
    
    \vspace{-1mm}
    \caption{\textit{Cluster parameters obtained from literature survey \cite{CG}. Column 2 and 3 represent the coordinates of the corresponding cluster, RA and Dec, respectively. Columns 4 and 5 represent the PM along RA and Dec, respectively. D is the distance to the cluster, r50 is the radius containing half of the cluster members, Rmax is the r50$\times$2 in arcminutes, and the last column represents the $log_{10}(age)$.}}
    
    \label{table_apx}
\end{table}

\begin{table}[h]
    \centering
    \renewcommand{\arraystretch}{1.5}
    \resizebox{17cm}{!}{
        \begin{tabular}{l |l ll  |l l l l  |l l l l}
            \hline
            Cluster name & \multicolumn{3}{c|}{$N_{star}$}& \multicolumn{4}{c|}{D(pc)}& \multicolumn{4}{c}{$log_{10}(age)$ (dex)}\\
            & This study &Cantat -Gaudin& Dias& This study& Cantat-Gaudin& Kharchenko&Dias& This study&Cantat-Gaudin& Kharchenko&Dias\\
            \hline
            Alessi 24         & 135&157 & 159& $500.160^{+0.970}_{-0.811}$  & 502  & ... &482& $7.795^{+0.001}_{-0.002}$ & 7.86  & ... &8.457\\
            Alessi Teutsch 12 & 22&41 & 43& $646.998^{+0.015}_{-0.020}$  & 641  & ... &605& $7.788^{+0.004}_{-0.007}$ & 7.94  & ... &7.935\\
            ASCC 58           & 123&133 & 134& $490.998^{+0.100}_{-0.185}$  & 491  &564 &477& $7.601^{+0.005}_{-0.006}$ & 7.72  &7.2 &7.775\\
            Alessi Teutsch 5  & 118&127 & 147& $910.018^{+0.039}_{-0.024}$  & 907  & ... &869& $6.886^{+0.001}_{-0.016}$ & 7.27  & ... &7.071\\
            ASCC 16           & 280&175 & 207& $334.002^{+0.010}_{-0.009}$  & 344  &397 & 348& $7.157^{+0.005}_{-0.020}$ & 7.13  &7 & 7.088\\
            ASCC 21           & 223&90 & 102& $350.006^{+0.010}_{-0.013}$  & 341  &379 &343& $7.021^{+0.011}_{-0.001}$ & 6.95  &7.11 &7.102\\
            Collinder 69      & 517&620 & 532& $418.243^{+0.047}_{-0.069}$  & 416  &411 &400& $7.011^{+0.016}_{-0.057}$ & 7.1  &6.76 &6.946\\
            Collinder 132     & 186&86 & 98& $628.000^{+0.031}_{-0.040}$  & 624  &330 & 648& $7.445^{+0.031}_{-0.013}$ & 7.39  &7.51 & 7.488\\
            IC 2395           & 329&291 & 269& $704.001^{+0.006}_{-0.008}$  & 702  &701 &710& $6.897^{+0.005}_{-0.001}$ & 7.31  &7.27 &7.149\\
            ASCC 127          & 113&117 & 113& $350.944^{+13.498}_{-3.142}$ & 376  &400 &365& $7.147^{+0.001}_{-0.000}$ & 7.26  &7.82 &7.496\\
            NGC 1980          & 249&121 & 106& $351.984^{+6.310}_{-4.238}$  & 377  &520 &316& $6.940^{+0.008}_{-0.008}$ & 7.12  &6.67 &6.97\\
            Roslund 5         & 151&151 & 158& $555.011^{+0.009}_{-0.017}$  & 546  &508 &536& $7.946^{+0.008}_{-0.003}$ & 7.99  &7.57 &8.373\\
            UBC 26            & 140&64 & ... & $578.684^{+0.020}_{-0.012}$  & 582  & ... & ... & $7.482^{+0.023}_{-0.028}$& 7.55  & ... & ...\\
            NGC 6405          & 641&573 & 492& $468.003^{+0.020}_{-0.022}$  & 459  &356 &453& $7.697^{+0.009}_{-0.002}$ & 7.54  &8.035 &7.894\\
            \hline

        \end{tabular}
    }
    \caption{\textit{Comparison of cluster parameters with previous studies such as \cite{CG, Kharchenko2016, Dias2021}}}
    \label{Table:Compare}
\end{table}

\begin{figure}
    \centering
    \includegraphics[width=0.75\linewidth]{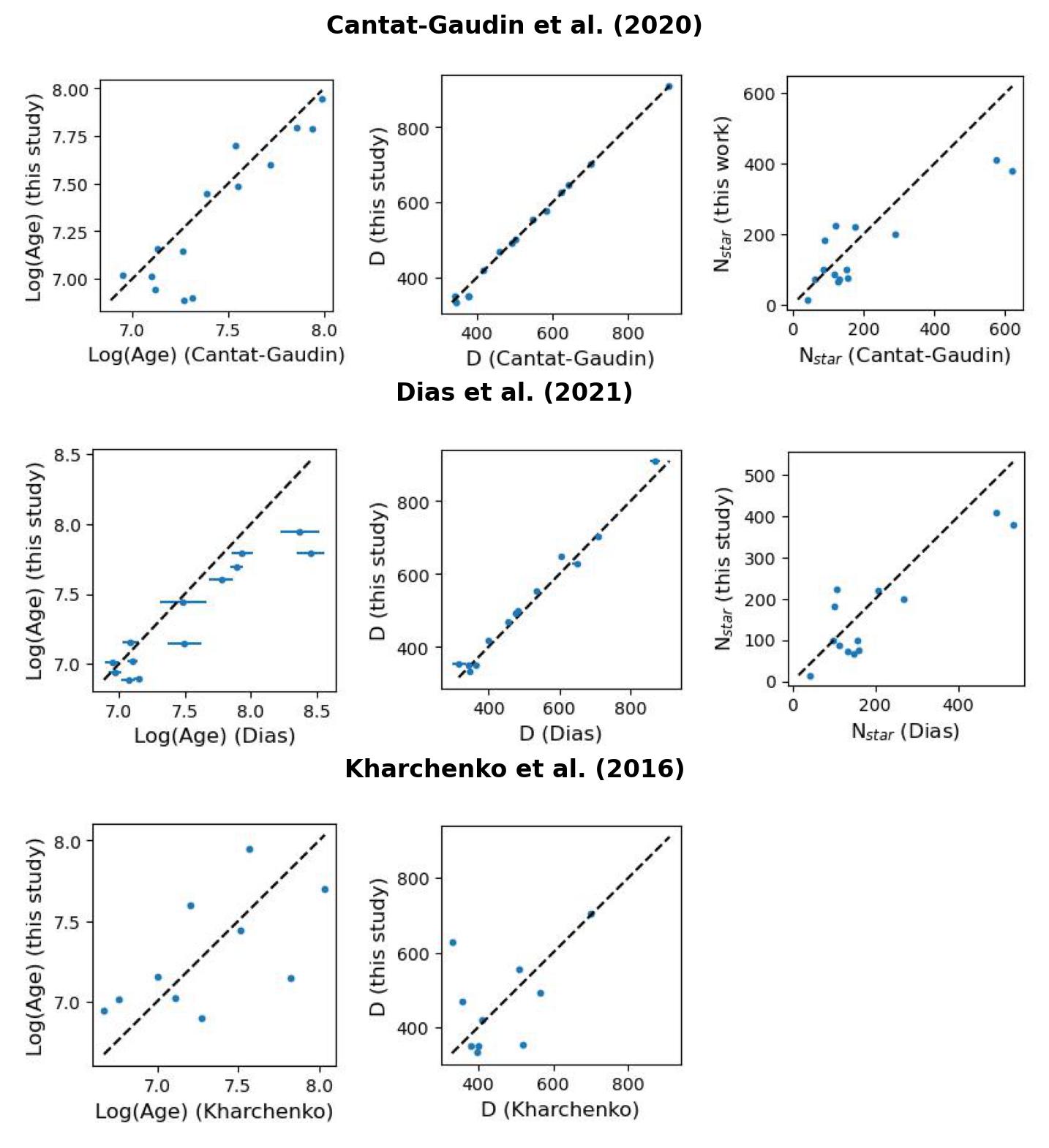}
    \caption{\textit{Comparison of cluster parameters with previous studies such as \cite{CG, Kharchenko2016, Dias2021}}}
    \label{fig:lit_compare}
\end{figure}

\clearpage
\twocolumn
\bibliographystyle{apj}

\bibliography{reference}
\balance


\end{document}